\documentclass[11pt,a4paper]{article}

\usepackage[utf8]{inputenc}
\usepackage{graphicx}
\usepackage{geometry}
\usepackage{amsmath}
\usepackage{booktabs}
\usepackage{hyperref}
\usepackage{authblk}
\usepackage{float}
\usepackage{caption}
\usepackage{cite}
\usepackage{authblk}
\usepackage{threeparttable}

\title{The Breakthrough of Sleep: A Contactless Approach for Accurate Sleep Stage Detection Using the Sleepal AI Lamp}

\author[1]{Zhuo Diao\thanks{Corresponding authors.}}
\author[1]{Yueting Li\protect\footnotemark[1]}
\author[1]{Jianpeng Wang}
\author[1]{Shengyu Guan}
\author[1]{Xinwei Wang}
\author[1]{Wenxiong Cui}
\author[1]{Xin Shi}
\author[1]{Tong Liu}
\author[2,3]{Kailai Sun}
\author[4]{Jingyu Wang}
\author[1]{Dian Fan}
\author[5]{Thomas Penzel}

\affil[1]{Hong Kong Xsmart Century Technology Co., Ltd., Hong Kong, China}
\affil[2]{Urban Mobility Lab, Massachusetts Institute of Technology, Cambridge, MA 02139, USA}
\affil[3]{Singapore-MIT Alliance for Research and Technology (SMART) Centre, Singapore 138602, Singapore}
\affil[4]{Department of Respiratory and Critical Care Medicine, Binzhou Medical University Hospital, Binzhou 256603, China}
\affil[5]{Interdisciplinary Center of Sleep Medicine, Charit\'{e} -- Universit\"{a}tsmedizin Berlin, 10117 Berlin, Germany}

\date{}

\begin{document}

\maketitle

\begingroup
\renewcommand\thefootnote{\textdagger} 
\footnotetext{A comprehensive technical description and evaluation of the methodological innovations regarding the sleep staging algorithm can be found in our companion paper co-authored with Prof. Thomas Penzel [Diao \textit{et al.}, 2026, submitted to IEEE Journal of Biomedical and Health Informatics (JBHI)].}
\addtocounter{footnote}{-1}
\endgroup

\begin{abstract}
Sleep staging is essential for the assessment of sleep quality and the diagnosis of sleep-related disorders.  Conventional polysomnography (PSG), while considered the gold standard, is intrusive, labor-intensive, and unsuitable for long-term monitoring.  This study evaluates the performance of the Sleepal AI Lamp,a contactless, radar-based consumer-grade sleep tracker, in comparison with gold-standard polysomnography (PSG), using a large-scale dataset comprising 1022 overnight recordings. We extract multi-scale respiratory and motion-related features from radar signals to train a frequency-augmented deep learning model. For the binary sleep-wake classification task, experimental results demonstrated that the model achieved an accuracy of 92.8\% alongside a macro-averaged F1 score of 0.895. For four-stage classification (wake, light NREM (N1 + N2), deep NREM (N3), REM), the model achieved an accuracy of 78.5\% with a Cohen’s kappa coefficient of 0.695 in healthy individuals and maintained a stable accuracy of 77.2\% with a kappa of 0.677 in a heterogeneous population including patients with varying severities of obstructive sleep apnea (OSA). These experimental results demonstrate that the sleep staging performance of the contactless Sleepal AI Lamp is in high agreement with expert-labeled PSG sleep stages. Our findings suggest that non-contact radar sensing, combined with advanced temporal modeling, can provide reliable sleep staging performance without requiring physical contact or wearable devices. Owing to its unobtrusive nature, ease of deployment, and robustness to long-term use, the contactless Sleepal AI Lamp shows strong potential for clinical screening, home-based sleep assessment, and continuous longitudinal sleep monitoring in real-world medical and healthcare applications.
\end{abstract}

\section{Introduction}
Sleep plays a critical role in human health, influencing cognitive function, emotional regulation, and cardiovascular and metabolic processes. Accurate sleep staging is fundamental for evaluating sleep architecture and diagnosing sleep disorders such as insomnia, sleep apnea, and narcolepsy. Polysomnography (PSG)\cite{venselrundoPolysomnography2019}, which records multiple physiological signals including electroencephalography (EEG), electrooculography (EOG), and electromyography (EMG), is widely regarded as the gold standard for sleep staging. However, PSG requires extensive sensor attachment, professional operation, and controlled laboratory environments, making it unsuitable for long-term, repeated, or large-scale sleep monitoring.

To overcome these limitations, recent research has explored alternative approaches for sleep monitoring, including wearable devices\cite{walchSleepStagePrediction2019,beattie2017estimation} and non-contact sensing technologies. Wearable devices have become increasingly popular due to their convenience and ability to support long-term monitoring outside laboratory environments. Commercial consumer wearables such as The Apple Watch, Oura Ring and Fitbit integrate multiple physiological sensors-including photoplethysmography (PPG), accelerometers, and temperature sensors to estimate sleep stages through machine learning algorithms.

While wearable sensors reduce system complexity compared to PSG, they still require continuous physical contact and may affect sleep comfort and compliance, particularly for elderly individuals and patients requiring long term monitoring. Non-contact sensing methods, such as radar-based systems, offer a promising solution by enabling unobtrusive measurement of physiological signals without disturbing the sleeper.

Radar-based sleep monitoring leverages the sensitivity of radio-frequency signals to detect subtle chest movements caused by respiration as well as larger body movements during sleep. These signals contain valuable information related to sleep dynamics and can be used to infer sleep stages \cite{rahmanDoppleSleepContactlessUnobtrusive2015,50198,yooUnsupervisedDetectionMultiple2023}. Kagawa et al. utilized 24 GHz Doppler radar to derive vital sign indices and applied statistical classifiers to distinguish sleep stages, demonstrating the potential of radar signals for non-contact sleep analysis\cite{kagawaSleepStageClassification2016}. Toften et al. validated a classification approach using respiration and body movement data that were collected by the commercial Somnofy IR-UWB radar system\cite{toftenValidationSleepStage2020a}. Lee et al. developed an attention-based Bi-LSTM model using 60 GHz FMCW radar signals to classify sleep stages in patients with obstructive sleep apnea, achieving an overall accuracy of 85.0\% for Wake, REM and NREM (N1 + N2 + N3) three-stage classification\cite{leeDevelopingDeepLearning2024}. However, many existing studies are limited by small datasets, simplified experimental settings, or lack of validation against large-scale PSG-labeled data, which restricts their clinical relevance and generalizability.

In this work, we present the validation results of the Sleepal AI Lamp, a radar-based contactless sleep monitoring product, on a large dataset comprising 1022 PSG-annotated overnight recordings. A deep sleep staging model was developed by jointly modeling respiratory characteristics and body movement patterns extracted from radar signals. The staging model achieved an overall accuracy of 92.8\%, a macro-averaged F1 score of 0.895, and a cohen’s kappa\cite{cohenCoefficientAgreementNominal1960a} of 0.791 for sleep-wake binary classification. For four-stage classification, the overall accuracy of the model was 77.2\% with a kappa of 0.677. 

Since a significant portion of the data was collected at clinical sleep centers, nearly 30\% of the 1,022 total recordings were obtained from patients with moderate-to-severe apnea ($\text{AHI} \ge 15$). Elevated AHI levels are typically associated with increased sleep fragmentation and frequent stage transitions, which pose a heightened challenge for automated sleep staging models. To assess the system's robustness, we evaluated the four-class staging performance across populations stratified by OSA severity. In the Normal group ($\text{AHI} \le 5$), the model achieved an accuracy of 78.5\% and a Cohen's Kappa of 0.695. For the Mild group ($5 < \text{AHI} \le 15$), the accuracy and Kappa were 75.3\% and 0.648, respectively. The Moderate group ($15 < \text{AHI} \le 30$) demonstrated stable performance with an accuracy of 77.9\% and a Kappa of 0.676. Notably, even in the Severe group ($\text{AHI} > 30$), the model maintained robust performance, attaining an accuracy of 74.3\% and a Kappa of 0.614.

In this study, we constructed and utilized a large-scale synchronized radar–PSG dataset comprising 1,022 overnight recordings for sleep staging—substantially larger than datasets reported in most prior studies. Experimental results demonstrate that non-contact radar sensing product can achieve sleep staging performance competitive with manual annotations derived from polysomnography. These findings provide robust evidence that non-contact radar-based sleep tracker represents a viable and scalable alternative for sleep monitoring, highlighting its strong potential for future integration into clinical practice and population-level sleep health management.

The remainder of this paper is organized as follows. Section 2 describes in detail the data acquisition and preprocessing mechanism, the extraction of respiratory, movement, and circadian features, as well as the staging model structure. Section 3 presents detailed experimental results, along with a comparison and analysis against PSG annotations. A discussion and the conclusions of this work are presented in Sections 4 and 5 respectively.

\section{Data and Methods}

\subsection{Data Acquisition Device}
\begin{figure}[H]
    \centering
    \includegraphics[width=0.8\textwidth]{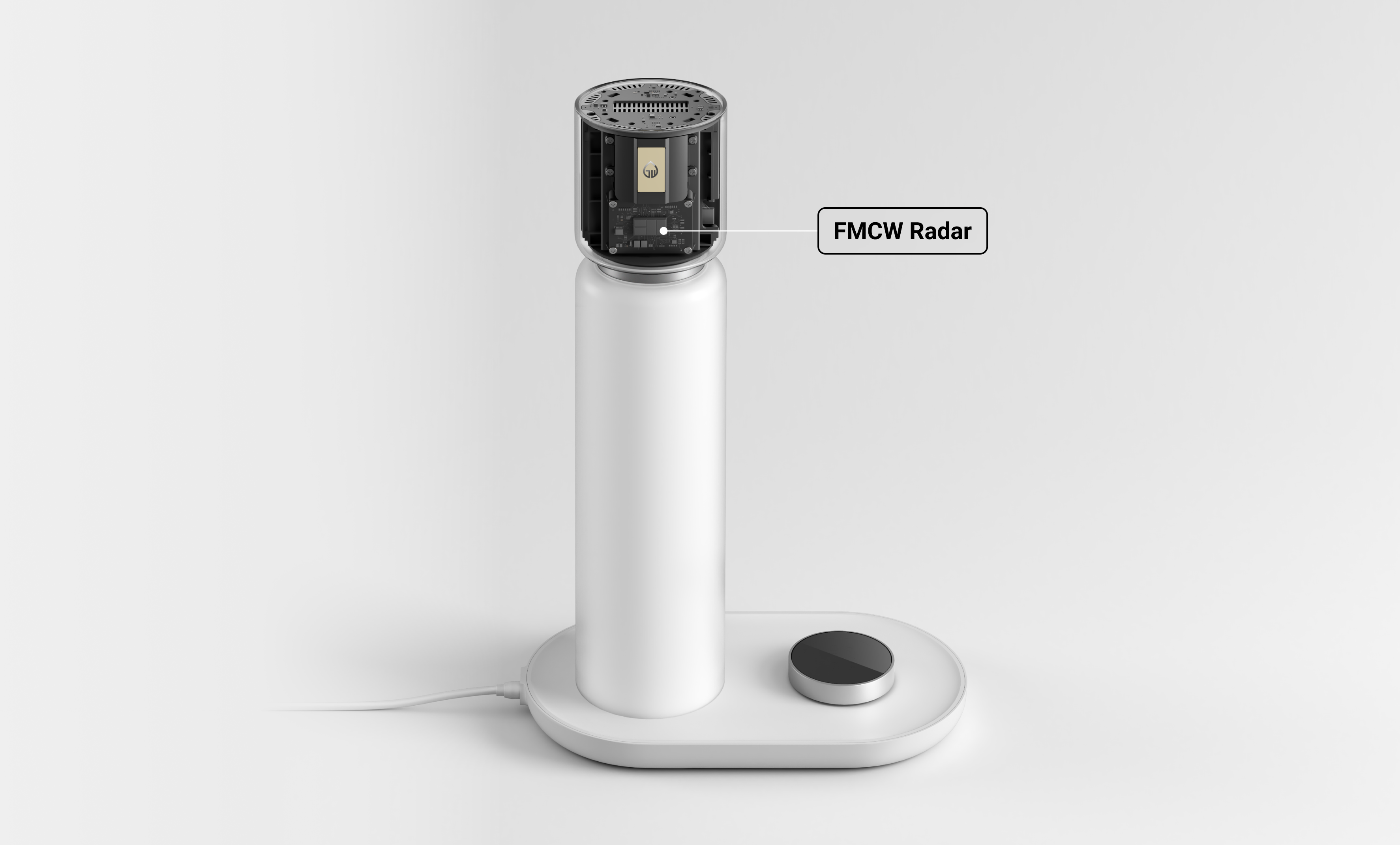} 
    \caption{Physical prototype of the Sleepal AI Lamp. The transparent black section at the top houses the core data acquisition module.}
    \label{fig:1}
\end{figure}

For model development and validation, sleep data were collected using a non-contact radar-based consumer product namely the Sleepal AI Lamp. As shown in Fig.\ref{fig:1}, Sleepal AI Lamp integrates a 60 GHz FMCW radar sensor designed for unobtrusive monitoring of human physiological signals during sleep. During data acquisition, a research version of Sleepal AI Lamp was placed at the bedside and oriented toward the subject's thoracic region, enabling continuous sensing without any physical contact or wearable attachments.

All measurements were conducted concurrently with a reference polysomnography (PSG) system, which provided expert-labeled sleep stage annotations. The radar and PSG recordings were temporally synchronized to ensure accurate alignment between radar-derived signals and PSG-based sleep stages for subsequent model training and evaluation.

\subsection{Datasets}
This multi-center study initially recruited a total of 1,117 participants, spanning a broad spectrum from healthy volunteers without known sleep complaints to patients referred for clinical evaluation of suspected sleep-disordered breathing. Data collection was conducted over a continuous seven-month period between 2025 and 2026. The primary inclusion criteria were: (1) aged 18 years or older; (2) willing and able to undergo standard overnight PSG and concurrent radar monitoring; and (3) a stable medical condition allowing for a baseline sleep assessment. Potential participants were excluded at the recruitment stage only if they had severe cognitive impairments that precluded informed consent.

Prior to dataset partitioning, a rigorous quality control and initial screening protocol was implemented to ensure the fidelity of the radar signals and the reliability of the reference data. First, sessions exhibiting severe signal corruption—such as prolonged PSG sensor detachment, excessive environmental interference, or significantly missing annotations—were systematically excluded. Second, we screened for specific confounding physiological and clinical cases. Subjects who experienced almost no sleep throughout the entire night were removed, as their recordings lacked meaningful sleep architecture necessary for effective model training. Furthermore, participants with conditions known to introduce severe continuous motion artifacts and disrupt radar-derived cardiopulmonary signals, such as severe restless legs syndrome, were also excluded from the study. 

After applying these exclusion criteria, the final dataset comprised 1022 valid overnight recordings. The detailed demographic characteristics and clinical sleep profiles of the included participants are summarized in Table \ref{tab:demographics}.

\begin{table}[htbp]
    \centering
    \begin{threeparttable}
        \caption{Demographic Characteristics and Sleep Data of Participants.}
        \label{tab:demographics}
        \begin{tabular}{lc}
            \toprule
            \textbf{Characteristics} & \textbf{Value (Mean $\pm$ S.D. or Count)} \\
            \midrule
            Gender (Male/Female) & 670 / 352 \\
            Age (years)          & 44.01 $\pm$ 15.88 \\
            BMI (kg/m$^2$)       & 26.80 $\pm$ 5.43 \\
            AHI (events/h)       & 14.01 $\pm$ 21.76 \\
            Time in Bed (TIB, min)          & 534.70 $\pm$ 61.82 \\
            Total Sleep Time (TST, min)     & 409.30 $\pm$ 61.98 \\
            Sleep Efficiency (SE, \%)       & 76.55 $\pm$ 13.84 \\
            Wake After Sleep Onset (WASO, min)\tnote{*} & 50.72 $\pm$ 46.53 \\
            Sleep Latency (SL, min)         & 56.99 $\pm$ 43.68 \\
            \bottomrule
        \end{tabular}
        \begin{tablenotes}
            \footnotesize
            \item[*] WASO was defined as the total duration of wakefulness occurring after sleep onset and before the final awakening, as determined by the gold-standard PSG annotations.
        \end{tablenotes}
    \end{threeparttable}
\end{table}

During each recording night, the Sleep AI Lamp was placed on a bedside table next to the bed under natural sleep conditions, while subjects simultaneously wore standard polysomnography (PSG) equipment for concurrent recording. PSG recordings were performed under standardized conditions using clinical-grade multi-channel monitoring configurations. This setup synchronously acquired neurophysiological signals for the determination of the sleep stage, as well as physiological parameters related to the evaluation of cardiopulmonary function and body movement. The Sleep AI Lamp and PSG systems were temporally synchronized to ensure precise alignment across different signal modalities.

Sleep stage annotations were based on the PSG recordings as the reference standard. All sleep stages were determined according to the scoring manual established by the American Academy of Sleep Medicine (AASM)\cite{berryAASMScoringManual2017a} and were scored by professionally trained and certified PSG technologists. These annotations served as the ground truth for subsequent model training and evaluation.

During data processing, continuous sleep recordings were divided into fixed 30-second epochs, with each epoch assigned a unique sleep stage label. Epochs with obvious signal loss or insufficient quality were discarded according to predefined quality control rules to ensure data reliability and consistency. Overall, the dataset encompasses a complete composition of sleep stages, exhibiting a naturally imbalanced distribution across different subjects and nights, which closely aligns with typical distributions observed in real-world clinical sleep recordings. Significant individual differences in sleep architecture and stage proportions were present, providing favorable conditions for evaluating the model's generalization capability in real-world populations.

The dataset was partitioned into a training set ($N = 762$) and an independent validation set ($N = 260$) using a subject-level stratified random sampling approach. To ensure a balanced representation of disease severity and physiological variability, the stratification was primarily based on the distribution of AHI severity (Normal, Mild, Moderate, and Severe) and gender. To prevent data leakage and ensure the reliability of the evaluation results, all data splitting for model training and evaluation was performed at the subject level; that is, all data from the same subject appeared exclusively in either the training set or the validation set, without overlapping. This splitting strategy ensures that the model performance evaluation reflects its actual generalization ability on unseen subjects.

To verify the balance of the split, we conducted statistical comparisons of the demographic and sleep-related characteristics between the training and validation cohorts. Independent t-tests for continuous variables and chi-square tests for categorical variables revealed no significant differences ($p > 0.05$) between the two sets in terms of age (training: $44.2 \pm 16.1$ vs. validation: $43.6 \pm 15.4$ years), gender distribution, and BMI. Most crucially, the clinical sleep profiles, including AHI and the percentage of each sleep stage (Wake, Light, Deep, and REM), showed consistent means and standard deviations across both groups. This statistical parity ensures that the validation set provides an unbiased and representative evaluation of the model's performance on a population with varied sleep architectures and pathological conditions.

\subsection{Radar Signal Processing}

\begin{figure}[!t]
    \centering
    \includegraphics[width=1.0\textwidth, height=0.5\textheight, keepaspectratio]{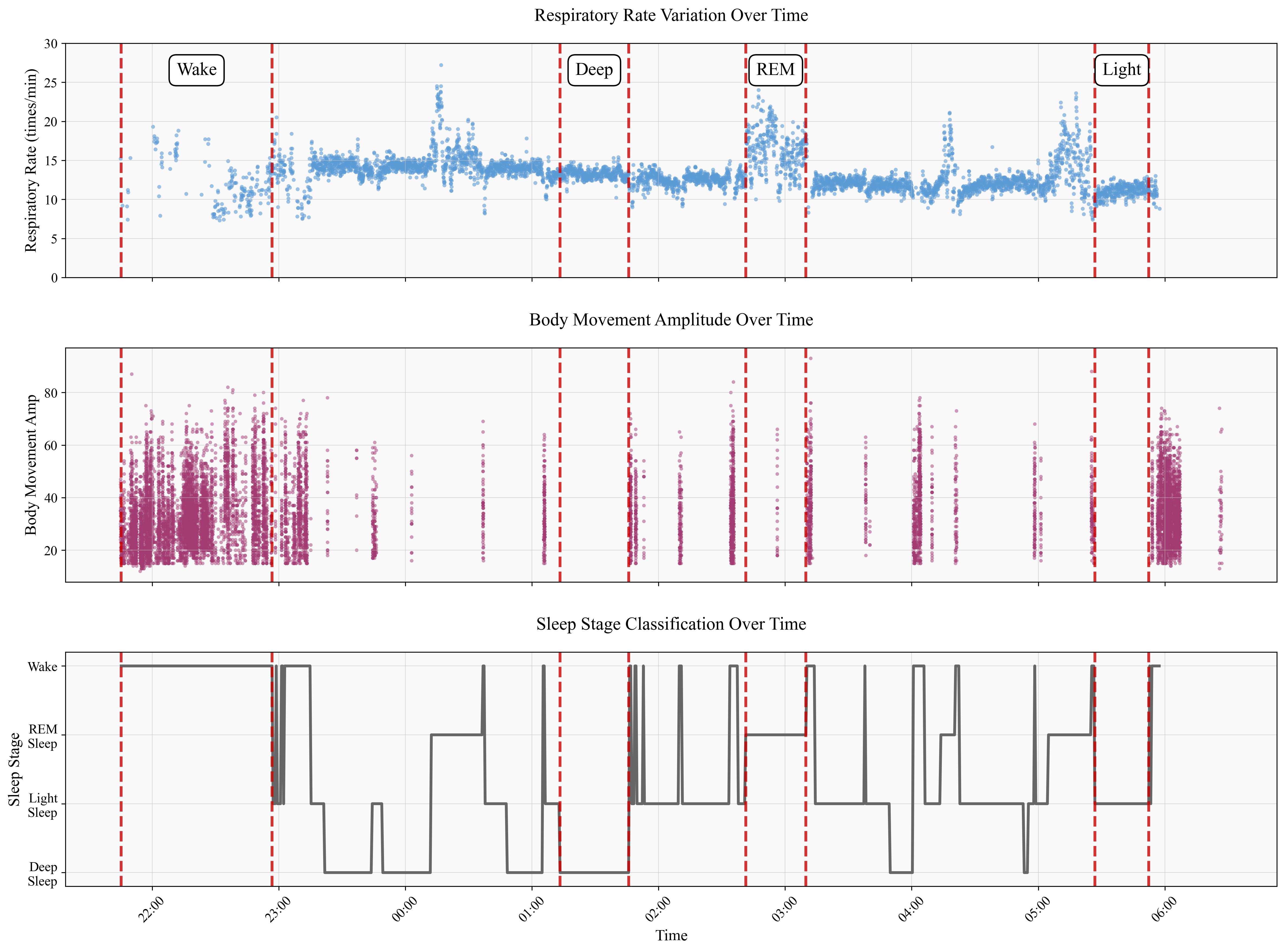}
    \caption{Time-synchronized visualization of radar-derived physiological signals and the corresponding expert-annotated sleep hypnogram over a full night. From top to bottom, the panels display the continuous respiratory rate variation, body movement amplitude, and the reference sleep stages. Dashed red vertical lines highlight representative segments of Wake, Deep, REM, and Light sleep, illustrating the distinct respiratory and motor patterns characteristic of each stage.}
    \label{fig:2}
\end{figure}

Respiratory monitoring was performed by extracting phase variations from the radar signals corresponding to periodic chest displacements. Phase information was extracted from the range bin associated with the participant's chest position. The extracted phase signal was subsequently unwrapped and bandpass filtered between 0.1 and 0.6 Hz to generate a respiratory waveform. The respiration rate was then estimated from the respiratory waveform using a combination of temporal peak detection and frequency-domain analysis.

During periods of significant movement or physical activity, respiratory rate estimation was suspended to avoid motion-induced artifacts. Instead, body movement was quantified using point cloud representations derived from FMCW radar processing. The amplitude and intensity of body movement were characterized based on the number, spatial distribution, and temporal dynamics of radar point cloud clusters generated by the participant's motion. 

Fig.\ref{fig:2} illustrates that respiration dynamics and body movement patterns varied systematically across different sleep stages. During wakefulness, frequent and pronounced body movements were observed, which frequently interrupted respiratory signal extraction and resulted in fragmented and highly variable respiration rate estimates. In light sleep stages, body movement activity decreased noticeably, and respiration rate showed a mild reduction accompanied by moderate variability. As sleep progressed into deep sleep, respiration became slower and more regular, exhibiting low variability, while body movements were minimal. In contrast, REM sleep was characterized by increased irregularity in respiratory patterns\cite{chungREMSleepEstimation2009a}, with a relatively elevated respiration rate compared to non-REM deep sleep. Body movements during REM sleep were present but less pronounced than during wakefulness, reflecting the physiological characteristics of this stage.

\subsection{Feature Extraction}
\textbf{1. Respiratory Features}

Respiration-related features were extracted at multiple temporal scales to capture both short-term variability and longer-term trends in breathing dynamics. First, statistical features—including trimmed mean (with 10\% of values removed from both tails), mean, standard deviation, root mean square of successive differences (RMSSD), and successive difference metrics—were computed within consecutive 30s epochs. Second, the same set of statistical features was further calculated over extended temporal windows, such as 90s, 150s, 270s, and longer durations, to characterize slower respiratory modulations across sleep stages.

Third, temporal similarity features were derived using dynamic time warping (DTW)\cite{sakoeDynamicProgrammingAlgorithm1978} to quantify the similarity between respiratory waveforms in adjacent 30s epochs, capturing transitional dynamics in respiration across epoch boundaries.

\textbf{2. Movement Features}

For each 30s epoch, movement-related features were derived from the temporal evolution of radar point cloud data. Specifically, the maximum number of consecutive frames containing detected movement, the total number of frames with movement activity, and the cumulative movement amplitude were computed to quantify movement duration, frequency, and intensity, respectively. These features enable discrimination between brief, involuntary movements—such as muscle twitches during REM sleep or short movements during non-REM sleep—and more sustained, volitional movements that typically occur during wakefulness or microarousal events.

\textbf{3. Chronobiological and Temporal Features}

Sleep macrostructure is profoundly governed by chronobiological rhythms and homeostatic sleep drive\cite{borbelyTwoProcessModel1982}. To incorporate these physiological priors into the computational framework, we engineered temporal embeddings to explicitly encode the global context of the sleep process. 

\begin{figure}[htb]
    \centering
    \includegraphics[width=1.0\textwidth]{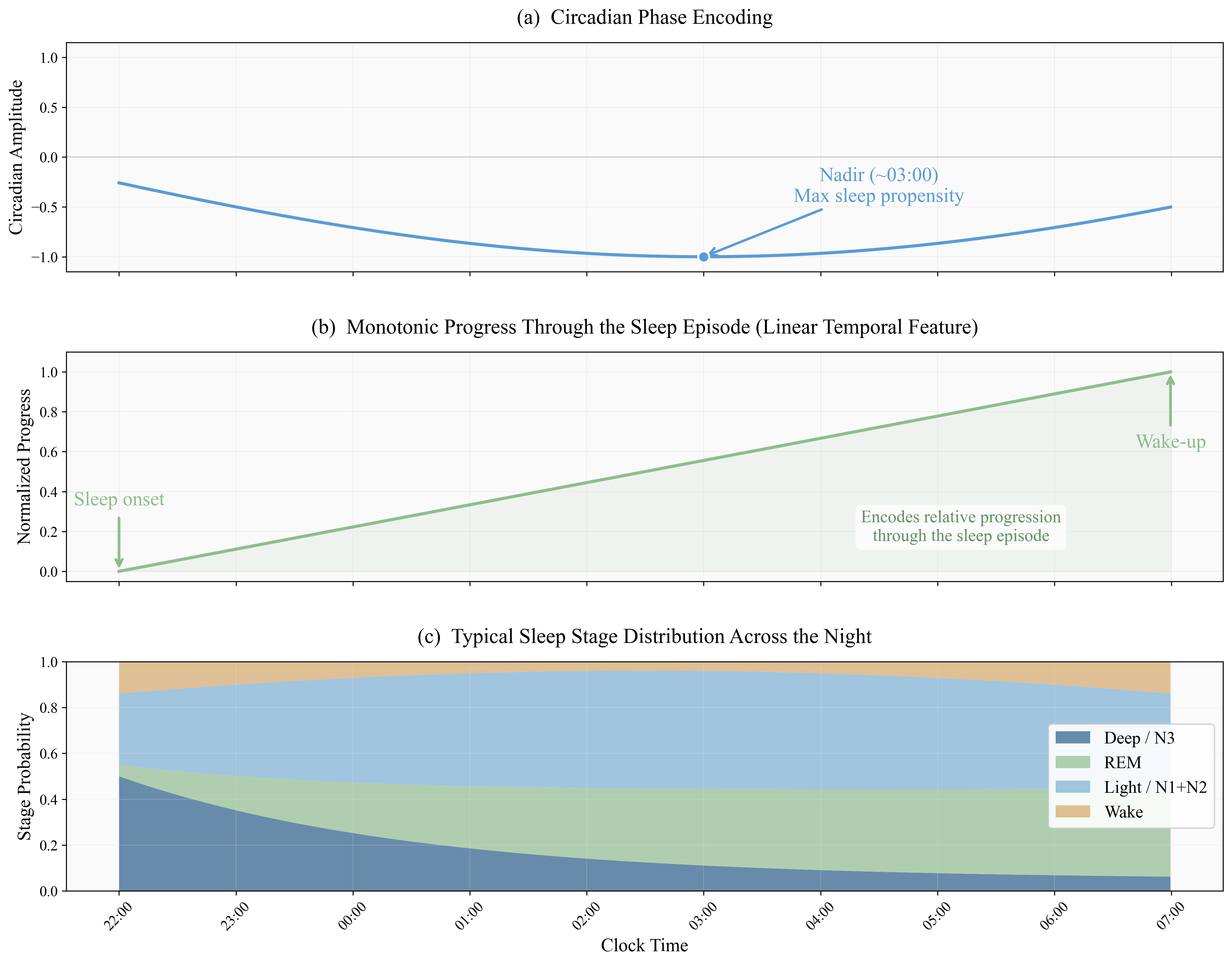} 
    \caption{Visualization of the chronobiological and temporal feature embeddings. \textbf{(a)} Circadian phase encoding using a phase-shifted cosine function. \textbf{(b)} Normalized linear temporal feature representing homeostatic sleep progression. \textbf{(c)} Typical probability distribution of sleep stages across the night.}
    \label{fig:3}
\end{figure}

Specifically, the circadian phase was mathematically approximated using a time-encoded cosine function. This continuous cyclical representation maps absolute clock time to a normalized amplitude, which is phase-shifted to reach its nadir (minimum) at approximately 03:00\cite{czeislerStabilityPrecisionNear24Hour1999}. This nadir physiologically aligns with the period of maximal sleep propensity and minimal core body temperature. By injecting this chronobiological prior, the system gains global contextual awareness of the typical diurnal sleep-wake cycle, as illustrated in Fig. \ref{fig:3}.

Furthermore, to capture the ultradian dynamics and the asymmetric evolution of sleep stages across the night, a normalized linear temporal feature was introduced. This component acts as a surrogate for the homeostatic sleep progression, reflecting the established polysomnographic consensus that slow-wave sleep (Deep/N3) predominantly occurs during the first half of the night, whereas rapid eye movement (REM) sleep episodes become progressively denser and longer toward the morning. The joint integration of these cyclical and linear components serves to explicitly model the temporal context of the night. Such explicit temporal modeling provides essential prior probabilities for sleep staging, effectively resolving physiological ambiguities based on the natural time-of-night distribution of different sleep stages.

\subsection{Temporal–Frequency Modeling Framework}

The proposed sleep staging framework employs a hierarchical deep learning architecture\textsuperscript{\textdagger} designed to capture both the sequential dynamics and spectral characteristics of nocturnal physiological signals. As the foundational backbone, a temporal branch comprising stacked bidirectional long short-term memory (BiLSTM)\cite{schusterBidirectionalRecurrentNeural1997} layers is utilized. This module systematically ingests the sequential multimodal features, effectively encoding both past and future contextual dependencies across consecutive epochs to model the continuous progression of sleep architecture.

To further capture the complex periodicities and physiological rhythms inherent in sleep stages, the representations produced by the BiLSTM are further enhanced in the frequency domain. Specifically, temporal features are transformed to extract informative spectral characteristics, enabling the model to better capture periodic patterns associated with sleep dynamics. By jointly modeling temporal dependencies and frequency-domain representations, the network benefits from complementary information across the two domains, allowing it to more effectively emphasize discriminative patterns related to subtle sleep stage transitions.

\begin{figure}[htbp]
    \centering
    \includegraphics[width=1.0\textwidth]{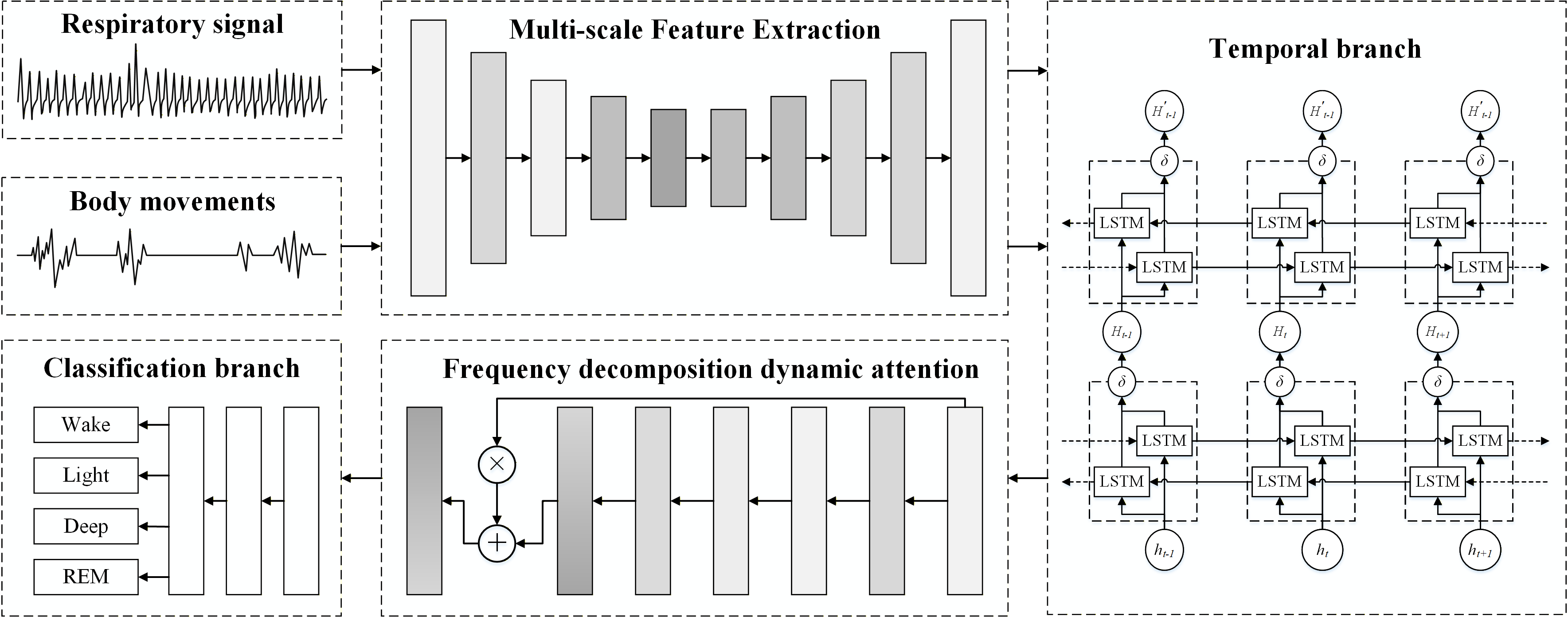} 
    \caption{Schematic overview of the proposed sleep staging architecture.}
    \label{fig:4}
\end{figure}

The refined, attention-weighted representations are ultimately fed into a robust classification branch. To ensure optimization stability and mitigate internal covariate shift, the features are first standardized via Layer Normalization (LayerNorm)\cite{baLayerNormalization2016a}. A linear projection layer, coupled with dropout regularization and an H-swish\cite{howardSearchingMobileNetV32019a} activation function, serves as the decision-making unit. Finally, a temperature-scaled softmax layer calibrates the outputs to yield the final probabilistic distribution over the target sleep stages. The comprehensive architecture of the model is illustrated in Fig. \ref{fig:4}.

\section{Results}

\subsection{Sleep-Wake Binary Classification Performance}

Evaluated on the validation set for the sleep-wake binary classification task, the proposed method demonstrated robust and balanced discriminative performance. At the epoch level, the model achieved an overall accuracy of 92.8\%, a macro-averaged F1 score of 0.895, and a Cohen's Kappa of 0.791. These metrics indicate a substantial agreement with the expert-labeled PSG annotations and a well-maintained balance between precision and recall, despite the inherent class imbalance between sleep and wakefulness across a full night.

\begin{figure}[!htbp]
    \centering
    \includegraphics[width=0.7\textwidth]{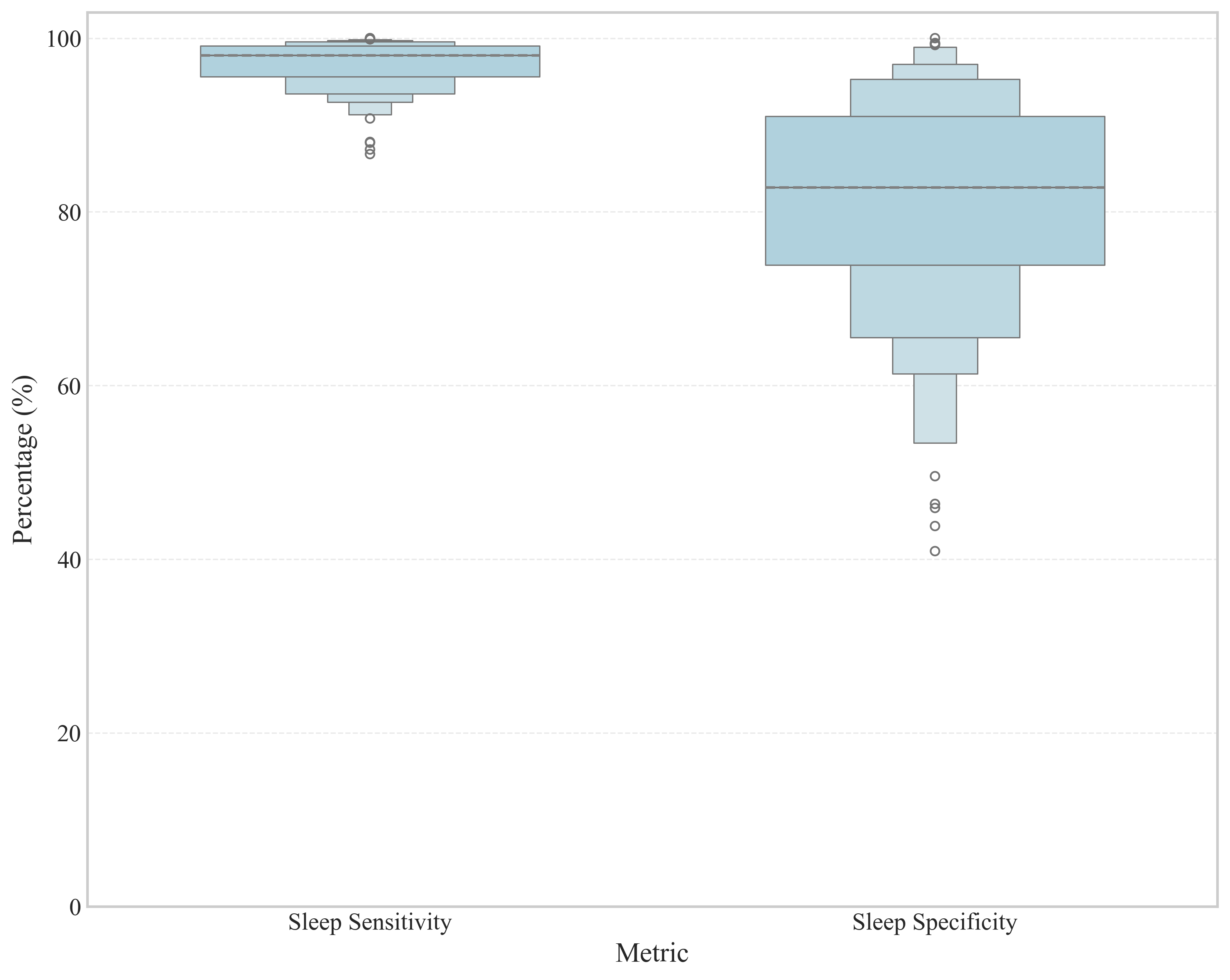}
    \caption{Letter-value plot visualizing the session-level performance for binary sleep-wake classification across the validation cohort. Sleep sensitivity (left) and sleep specificity (right) are computed for each subject. In this visualization, the center line denotes the median, and the primary boxes immediately above and below represent the interquartile range (IQR). Successive boxes partition the remaining percentile ranges in half at each step, providing a detailed view of the distribution tails, with outliers represented as individual points.}
    \label{fig:5}
\end{figure}

To assess the algorithm's consistency across the population, the performance was further evaluated at the session level. As illustrated in the letter-value plot (Fig. \ref{fig:5}), the system demonstrated a high and stable Sleep Sensitivity, achieving an average of $0.971 \pm 0.027$ (median: 0.980). The extremely narrow distribution of sensitivity scores suggests that the Sleepal AI Lamp is highly robust in capturing sleep epochs, exhibiting a clear "ceiling effect" across the entire validation cohort.

In contrast, Sleep Specificity (which reflects the accuracy of wakefulness detection) yielded a mean value of $0.812 \pm 0.127$ (median: 0.828). As shown in the plot, specificity exhibited a significantly broader distribution compared to sensitivity. This increased variance, including the presence of low-performing outliers in the distribution's tail, underscores the primary technical challenge of non-contact monitoring: the physiological ambiguity of "quiet wakefulness." During periods when subjects remain physically motionless despite being awake, their radar-derived cardiopulmonary and gross motor profiles closely mimic the signatures of N1 or light sleep, leading to a higher rate of false-positive sleep detections.

Analysis of these misclassifications reveals that while true sleep is rarely missed (high sensitivity), the system's errors are predominantly concentrated in mislabeling quiet wakefulness as sleep (reduced specificity). This trade-off is common in radar-based systems that prioritize the continuity of sleep architecture. Nevertheless, the high median performance remains within a clinically acceptable range, providing a reliable and robust front-end filter for subsequent four-class sleep staging tasks.

\subsection{Estimation Accuracy of Sleep Boundaries}

To evaluate the temporal precision of the proposed system in detecting sleep transitions, we analyzed the discrepancies between the model-predicted and PSG-annotated sleep onset and offset times. Discrepancies were calculated as the predicted time minus the true time (i.e., Prediction $-$ True) and aggregated at the session level.

\begin{figure}[!htbp]
    \centering
    \includegraphics[width=1.0\textwidth]{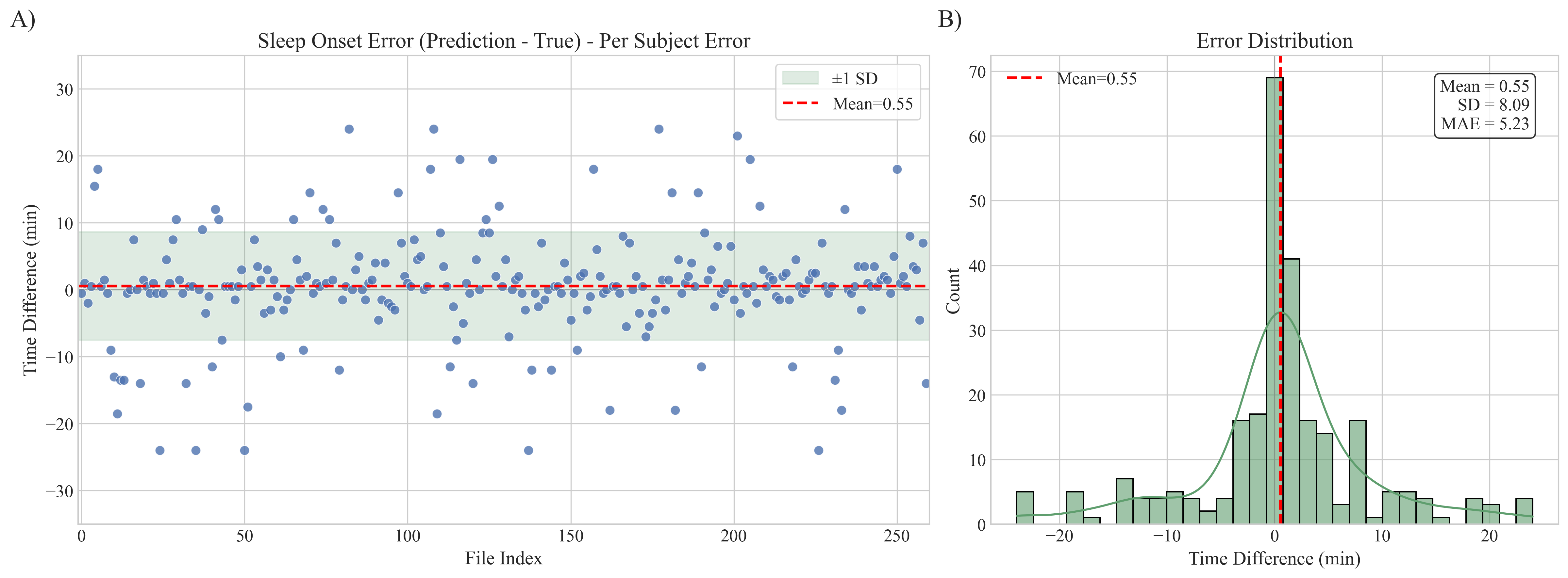}
    \caption{Evaluation of sleep onset prediction. (A) Scatter plot of per-subject time difference (Prediction - True) with the $\pm 1$ standard deviation (SD) band shaded. (B) Histogram and density curve of the error distribution.}
    \label{fig:6}
\end{figure}

As illustrated in Fig. \ref{fig:6}, the model demonstrated high consistency in predicting sleep onset. The mean temporal bias was merely 0.55 minutes, with its 95\% confidence interval (CI) of $[-0.44, 1.53]$ minutes securely covering zero, indicating an absence of systematic early or late predictions. The mean absolute error (MAE) stood at 5.23 minutes. As shown in the error distribution (Fig. \ref{fig:6}B), the errors are tightly centered around zero but exhibit a slight heavy-tailed distribution. This variance physiologically reflects the progressive and continuous nature of the wake-to-sleep transition, which inherently introduces boundary ambiguity and inter-rater variability even in manual PSG scoring.

\begin{figure}[htbp]
    \centering
    \includegraphics[width=1.0\textwidth]{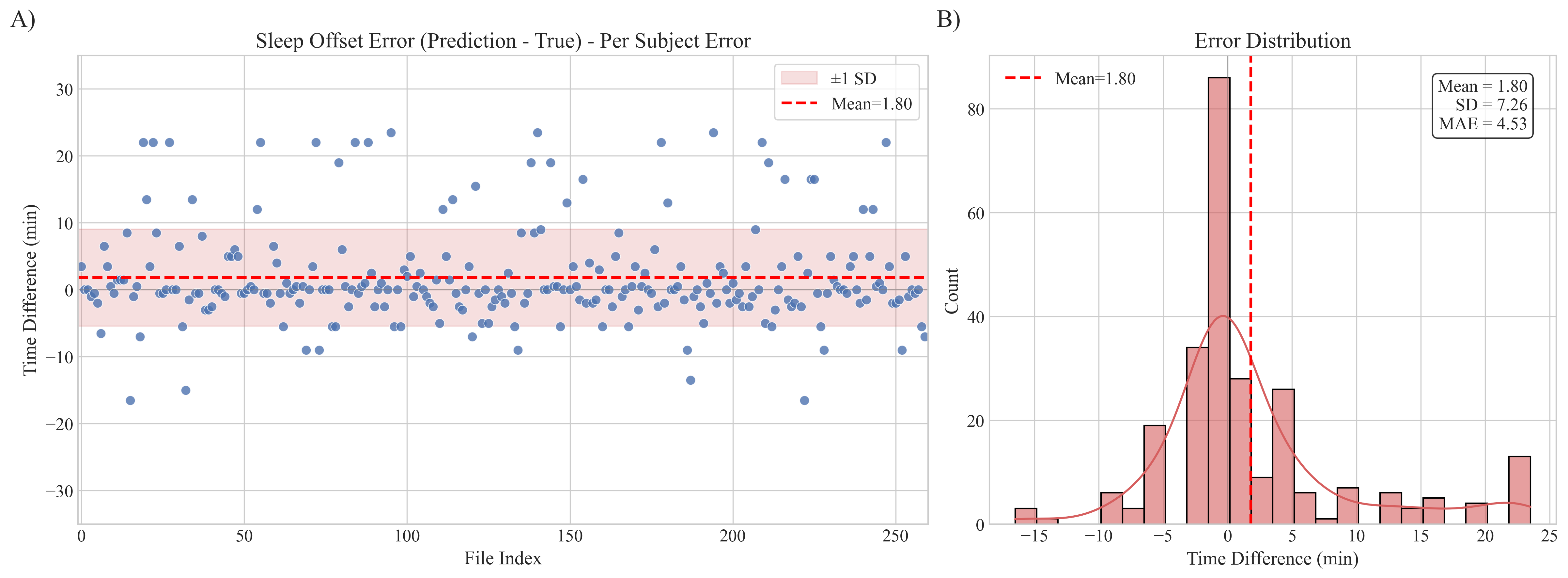}
    \caption{Evaluation of sleep offset prediction. (A) Scatter plot of per-subject time difference (Prediction - True) with the $\pm 1$ standard deviation (SD) band shaded. (B) Histogram and density curve of the error distribution.}
    \label{fig:7}
\end{figure}

Regarding sleep offset, depicted in Fig. \ref{fig:7}, the model exhibited a marginal positive bias. Specifically, the predicted sleep offset was, on average, 1.80 minutes later than the gold standard (95\% CI: $[0.92, 2.68]$ minutes). Although this delay presents a statistical bias, its magnitude is clinically negligible, maintaining a low MAE of 4.53 minutes. In non-contact sleep monitoring, such minor overestimations of sleep boundaries at the end of the night are common. They primarily stem from brief periods of quiet wakefulness (e.g., lying still in bed upon awakening), which project cardiopulmonary and gross motor profiles highly analogous to those of actual sleep.

From the perspective of session-level distribution, the scatter plots for both onset and offset errors (Fig. \ref{fig:6}A and Fig. \ref{fig:7}A) reveal no sample-dependent drift across the 260 validation nights. The vast majority of prediction errors fall strictly within the $\pm 1$ standard deviation (SD) shaded region, underscoring the model's stable generalization across diverse individuals. The sparse outliers contributing to the distribution's tail largely represent subject-specific atypical sleep behaviors rather than systemic algorithmic failures.

Ultimately, the proposed non-contact framework achieves high temporal fidelity with the gold standard in anchoring sleep boundaries. This precision ensures that derived clinical macro-parameters—such as sleep onset latency (SOL), sleep efficiency (SE), and wake after sleep onset (WASO)—are calculated upon highly reliable temporal demarcations.

\subsection{Four-Class Sleep Staging Performance and Duration Agreement}

Evaluated on the independent validation cohort, the proposed framework demonstrated robust and clinically viable performance for the four-class sleep staging task. At the epoch level, the model achieved a global accuracy of 77.2\%, a macro-averaged F1 score of 0.768, and a Cohen's Kappa coefficient of 0.677, firmly placing it within the domain of "substantial agreement" for automated sleep staging.

\begin{figure}[!htbp]
    \centering
    \includegraphics[width=0.55\textwidth]{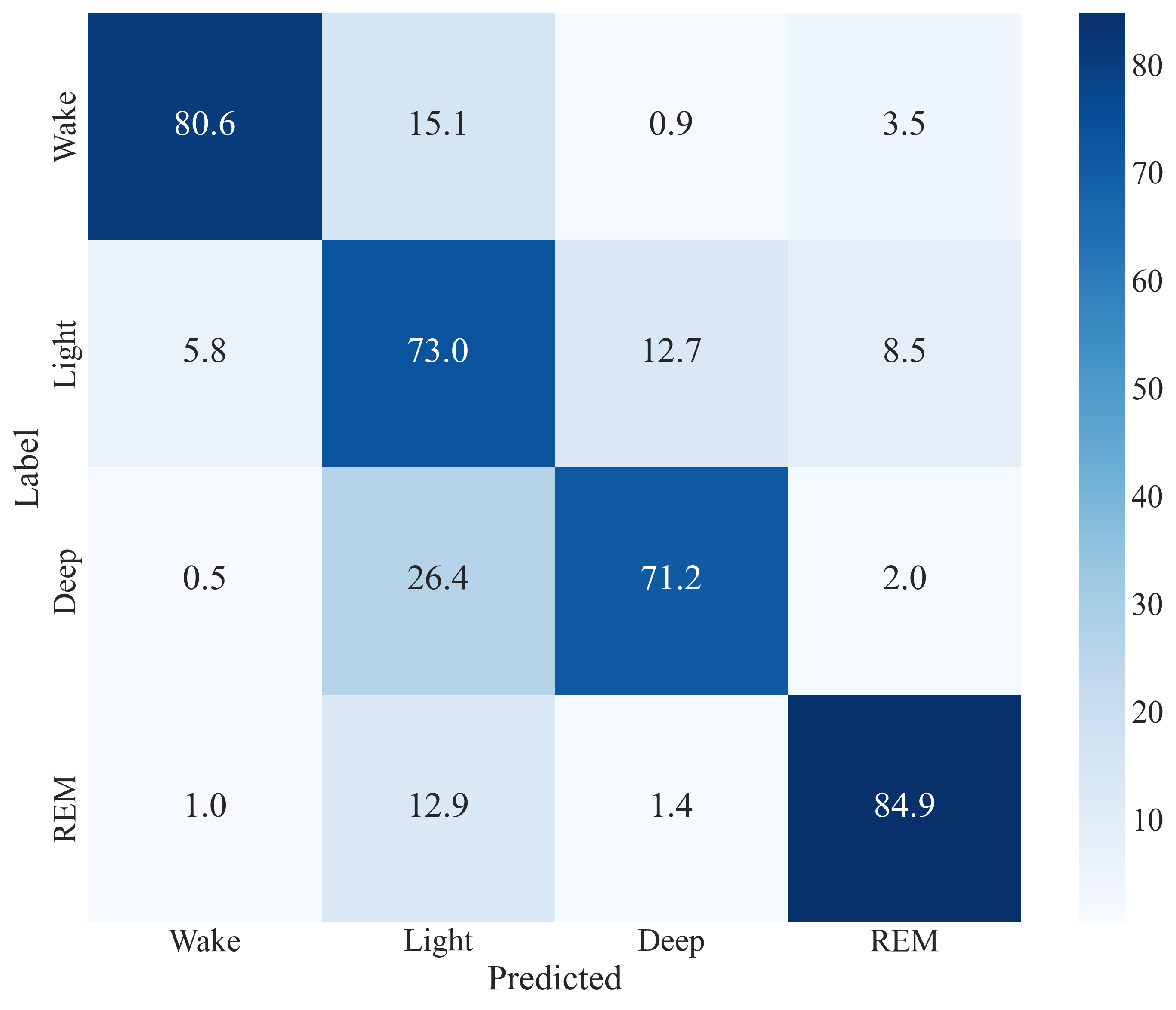} 
    \caption{Confusion matrix of the four-class sleep staging across the independent validation set.Each matrix is row-normalized to represent the recall (sensitivity) of each sleep stage, with values indicating the percentage of 30-s epochs.}
    \label{fig:8}
\end{figure}

\begin{figure*}[!b]
    \centering
    \includegraphics[width=0.9\textwidth]{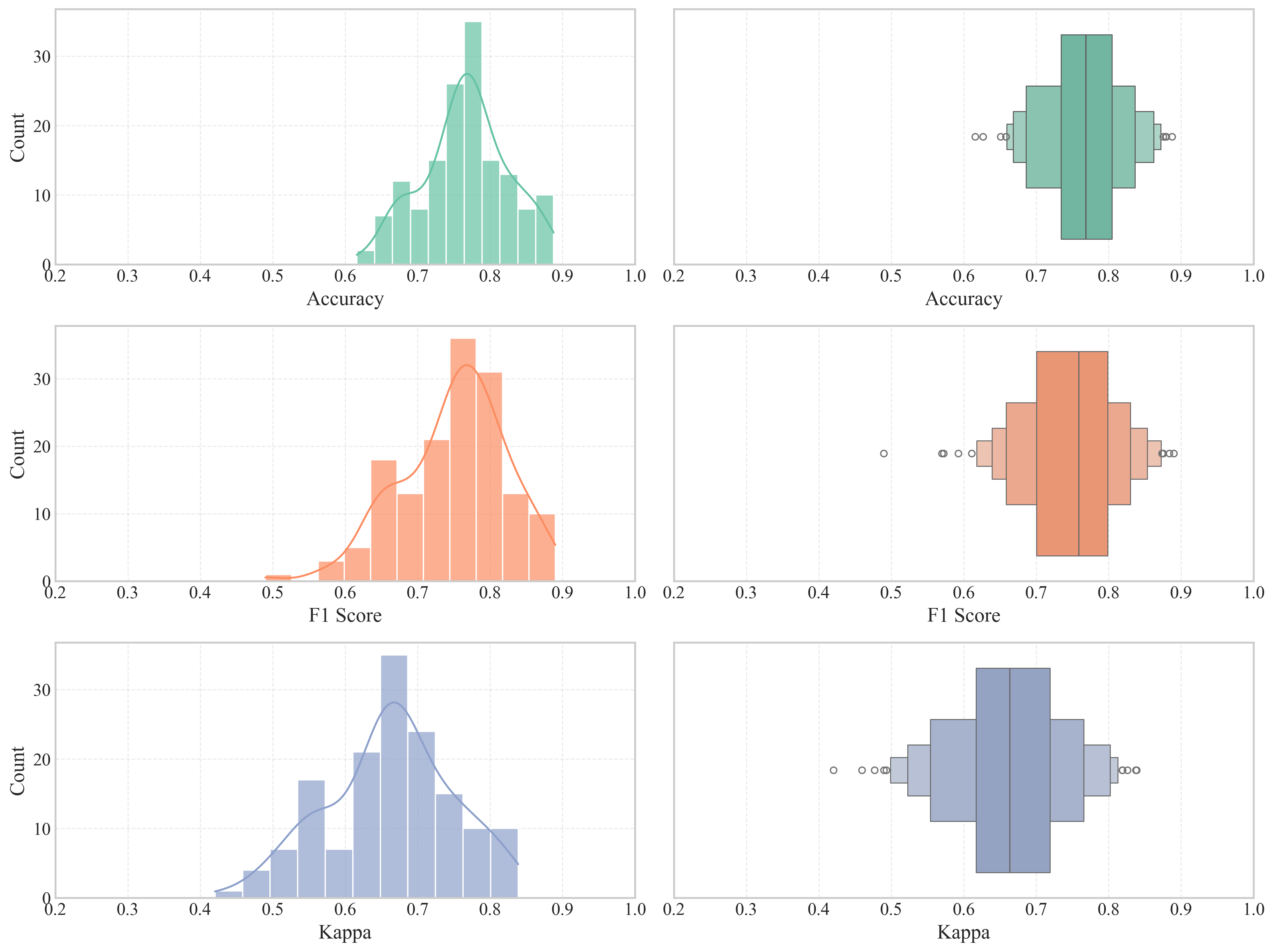}
    \caption{Subject-level performance distributions for Accuracy, F1 Score, and Cohen’s Kappa across the validation cohort ($N=260$). Left panels show histograms with kernel density estimate (KDE) curves, while right panels present horizontal letter-value plots (boxenplots) visualizing distribution density and tails.}
    \label{fig:9}
\end{figure*}

To further unpack this global performance, a detailed examination of the confusion matrix (Fig. \ref{fig:8}) uncovers the model's stage-specific discriminative characteristics. The REM stage was identified with high precision, achieving a recall rate of 84.9\% (F1 = 0.802). This indicates that the algorithm successfully captured the unique respiratory modulations inherent to REM sleep, resisting pathological interferences. The Wake stage also demonstrated excellent boundary discrimination, yielding a recall of 80.6\% and the highest F1 score (0.839) among all classes. For non-rapid eye movement (NREM) stages, the recall for Light Sleep (N1+N2) was 73.0\% (F1 = 0.740), while Deep Sleep (N3) achieved a recall of 71.2\% (F1 = 0.691). A predictable degree of cross-misclassification exists between these two stages, which inherently aligns with the physiological continuum of the N2-to-N3 transition. Given that human inter-rater reliability between standard PSG technicians at the N2/N3 boundary typically hovers around 70\%--80\%, the model's discriminative performance on NREM stages closely approaches the theoretical upper limit of manual scoring agreement.

Beyond the aggregate epoch-level results, the model exhibited remarkable consistency across individual subjects, as depicted in the performance distributions (Fig. \ref{fig:9}). The subject-level accuracy was $0.772 \pm 0.059$, the macro-averaged F1 score was $0.770 \pm 0.072$, and the Kappa coefficient was $0.667 \pm 0.087$. The distributions of these three core metrics (Fig. \ref{fig:9}, left) show pronounced central tendencies. Furthermore, the box-and-whisker plots (Fig. \ref{fig:9}, right) reveal compact interquartile ranges with negligible severe outliers, statistically confirming the model's reliable generalization across a highly heterogeneous population with diverse sleep architectures. In fact, over 75\% of the evaluated subjects attained an accuracy above 0.734, an F1 score above 0.701, and a Kappa above 0.617.

\begin{figure}[!b]
    \centering
    \includegraphics[width=1.0\textwidth]{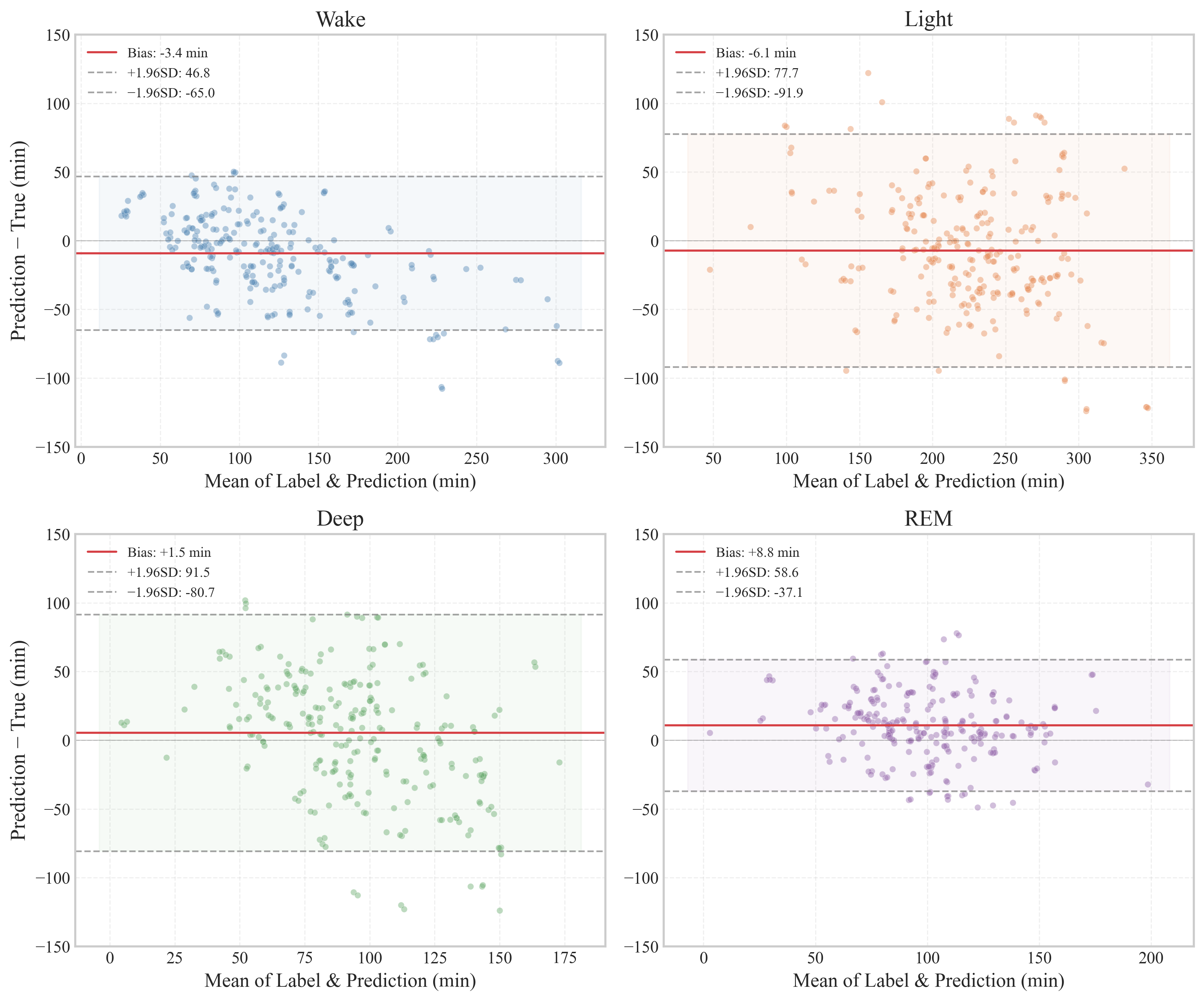}
    \caption{Bland-Altman plots assessing the agreement between predicted and ground-truth sleep stage durations (Wake, Light, Deep, and REM).}
    \label{fig:10}
\end{figure}

\begin{figure}[!htbp]
    \centering
    \includegraphics[width=0.95\textwidth, keepaspectratio]{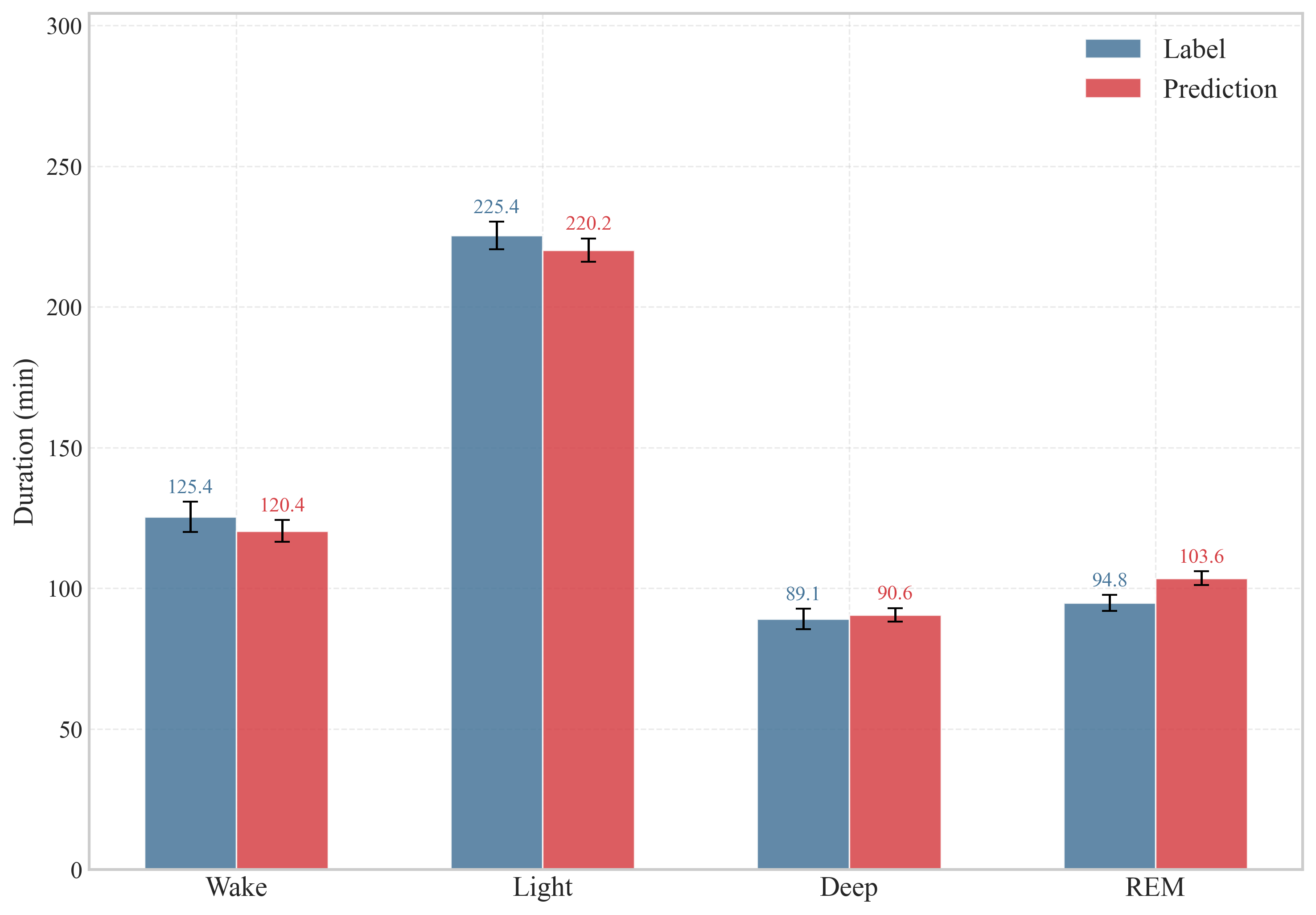} 
    \caption{Comparison of mean sleep stage durations between PSG annotations (Label) and model predictions.}
    \label{fig:11}
\end{figure}

Beyond epoch-level classification, precisely estimating the total duration of each sleep stage is critical for clinical assessments. To this end, we compared the model-predicted stage durations against the PSG gold standard using Bland-Altman\cite{martinblandSTATISTICALMETHODSASSESSING1986} agreement analysis (Fig. \ref{fig:10}), with the aggregated duration statistics detailed in Table \ref{tab:duration} and Fig. \ref{fig:11}. 

The mean bias across all four stages was stringently controlled within 9 minutes, with the vast majority of individual variations falling within the 95\% Limits of Agreement (LoA). As detailed in Table \ref{tab:duration}, the model exhibited a marginal underestimation of wake duration (Bias = $-5.0$ min). This negative bias accounts for merely 4.0\% of the true mean wake duration (125.4 min). Notably, the Bland-Altman scatter plots (Fig. \ref{fig:10}) revealed no evidence of proportional bias—where errors might scale with duration magnitude—confirming consistent estimation across varied sleep profiles. 

\begin{table}[!htbp]
    \centering
    \caption{Comparison of sleep stage durations (minutes) between label and model predictions.}
    \vspace{2mm}
    \label{tab:duration}
    \begin{tabular}{lccc}
        \toprule
        Stage & True (mean $\pm$ std) & Pred (mean $\pm$ std) & Diff (mean $\pm$ std) \\
        \midrule
        Wake  & 125.4 $\pm$ 66.7 & 120.4 $\pm$ 48.6 & $-5.0 \pm 30.6$ \\
        Light & 225.4 $\pm$ 60.8 & 220.2 $\pm$ 50.7 & $-5.2 \pm 44.7$ \\
        Deep  &  89.1 $\pm$ 44.4 &  90.6 $\pm$ 29.1 & $+1.5 \pm 44.1$ \\
        REM   &  94.8 $\pm$ 34.7 & 103.6 $\pm$ 29.9 & $+8.8 \pm 24.5$ \\
        \bottomrule
    \end{tabular}
\end{table}

Light sleep exhibited the smallest relative discrepancy across all stages (Bias = $-5.2$ min, representing 2.3\% of the true mean). The near-zero systematic bias proves that the model accurately delineates the dominant sleep stage without systemic over- or under-estimation. The comparatively wider absolute LoA interval for Light sleep is primarily attributable to the substantial natural inter-individual variability in this stage (True SD = 60.8 min).

Deep sleep prediction yielded a negligible overestimation (Bias = $+1.5$ min). This minimal discrepancy accounts for only 1.7\% of the true mean (89.1 min) and boasts the highest LoA coverage (96.0\%). This finding highlights that, despite the localized difficulties in pinpointing the exact N2/N3 epoch boundary, the model achieves a nearly unbiased estimation of cumulative overnight Deep sleep.

Finally, the REM stage displayed the largest absolute bias ($+8.8$ min), accounting for 9.3\% of the true mean duration. The model tended to slightly overestimate REM duration (Predicted 103.6 min vs. True 94.8 min). However, it is clinically significant that the REM stage also presented the narrowest LoA interval (width of 95.7 min) and the smallest standard deviation of differences (SD = 24.5 min). This indicates that while a minor systemic offset is present, the model's REM detection is remarkably uniform and stable across subjects.

\subsection{Staging Accuracy and Robustness Across Different AHI Populations}

\subsubsection{Overall Classification Performance and Inter-Group Differences}
To evaluate the impact of obstructive sleep apnea (OSA) severity on the proposed staging algorithm, subjects were stratified into four groups based on their Apnea-Hypopnea Index (AHI): Normal (AHI $\le 5$), Mild ($5 < \text{AHI} \le 15$), Moderate ($15 < \text{AHI} \le 30$), and Severe ($\text{AHI} > 30$). Because the model relies exclusively on cardiopulmonary and gross motor dynamics, the frequent respiratory events and associated micro-arousals inherent to pathological states inevitably disrupt the physiological baseline. 

\begin{table}[htbp]
    \centering
    \caption{Subject-level classification performance across different AHI severity groups (mean $\pm$ std).}
    \label{tab:ahi_metrics}
    \begin{tabular}{lcccc}
        \toprule
        AHI Group & $N$ & Accuracy (\%) & F1 (macro) & Cohen's Kappa \\
        \midrule
        Normal (AHI $\le$ 5)                & 124 & 78.5 $\pm$ 6.1 & 0.777 $\pm$ 0.079 & 0.695 $\pm$ 0.091 \\
        Mild ($5 < \text{AHI} \le 15$)      & 55  & 75.3 $\pm$ 5.4 & 0.743 $\pm$ 0.057 & 0.648 $\pm$ 0.072 \\
        Moderate ($15 < \text{AHI} \le 30$) & 38  & 77.9 $\pm$ 5.9 & 0.767 $\pm$ 0.063 & 0.676 $\pm$ 0.087 \\
        Severe ($\text{AHI} > 30$)          & 43  & 74.3 $\pm$ 7.6 & 0.711 $\pm$ 0.122 & 0.614 $\pm$ 0.137 \\
        \bottomrule
    \end{tabular}
\end{table}

As detailed in Table \ref{tab:ahi_metrics}, the model's overall predictive performance exhibited a foreseeable, yet remarkably well-controlled, decline as OSA severity increased. The Normal group demonstrated the highest staging performance, yielding an accuracy of $78.5\% \pm 6.1\%$ and a Cohen's Kappa of $0.695 \pm 0.091$. As the AHI increased, the metrics showed slight fluctuations. The performance of the Moderate group (Accuracy $77.9\%$, Kappa $0.676$) was slightly lower than that of the Normal group and slightly higher than that of the Mild group, a variance likely attributable to the specific data distribution characteristics of its smaller sample size. In the Severe group ($\text{AHI} > 30$), extreme sleep fragmentation caused by repetitive apneas\cite{bianchiObstructiveSleepApnea2010a} inevitably led to the lowest mean accuracy ($74.3\% \pm 7.6\%$). The increased standard deviation in this group reflects the magnified inter-subject variability in breathing patterns among severe patients. Nevertheless, it is highly clinically relevant that its Kappa coefficient ($0.614 \pm 0.137$) remained firmly above 0.6, sustaining a statistically defined ``substantial agreement'' despite profound architectural sleep disruptions.

\subsubsection{Stage-Specific Resilience and Confusion Matrix Analysis}

\begin{figure}[!b]
    \centering
    \includegraphics[width=1.0\textwidth, height=0.7\textheight, keepaspectratio]{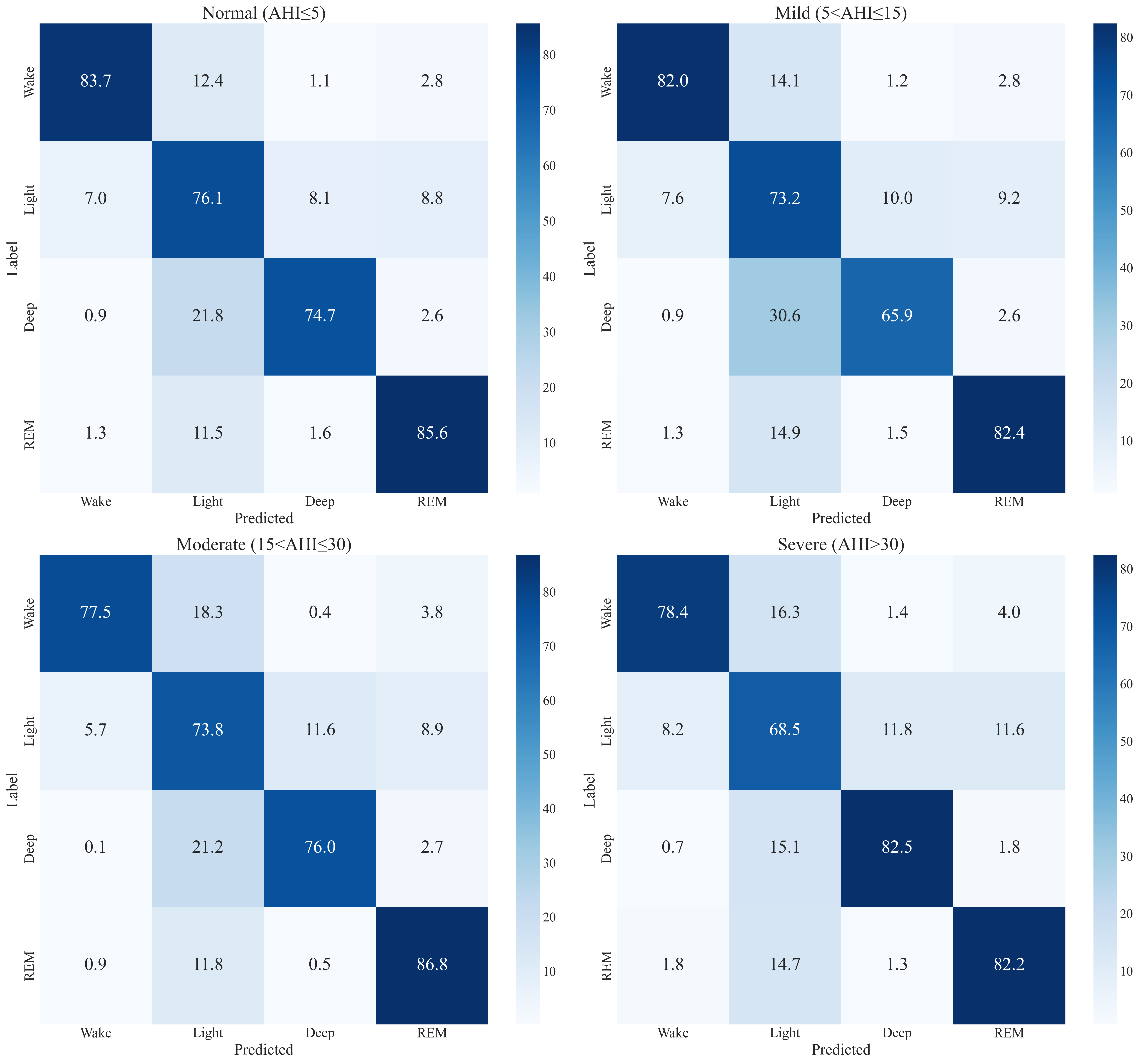}
    \caption{Normalized confusion matrices for the four-stage sleep classification stratified by Obstructive Sleep Apnea (OSA) severity. The four groups are defined by the Apnea-Hypopnea Index (AHI): Normal, Mild, Moderate, and Severe. Each matrix is row-normalized to represent the recall (sensitivity) of each sleep stage, with values indicating the percentage of 30-s epochs.}
    \label{fig:12}
\end{figure}

A granular analysis of the group-specific confusion matrices (Fig. \ref{fig:12}) elucidates how the algorithm resolves pathological respiration-movement features across distinct sleep stages.

Regarding the Wake state, the recognition rate slightly decreased from 83.7\% in the Normal group to 77.2\% in the Severe group. Wakefulness is primarily characterized by frequent body movements and highly irregular respiration. In severe OSA patients, however, approximately 16.0\% of true Wake epochs were misclassified as Light sleep (Fig. \ref{fig:12}, bottom right). From a signal perspective, this phenomenon is largely attributable to brief micro-arousals terminating respiratory events. These transient bursts of movement and compensatory hyperventilation are short-lived and often intermingle with the unstable physiological rhythms typical of Light sleep, leading to algorithmic confusion.

Light sleep proved to be the stage most susceptible to AHI-induced interference, with its recall dropping from 76.1\% (Normal) to 68.5\% (Severe). Physiologically, Light sleep marks a transition where respiration should gradually stabilize and body movements decrease. Conversely, in high-AHI patients, this stage is fraught with obstructive apneas and subsequent compensatory ventilations, causing violent fluctuations in respiratory frequency alongside intermittent tossing. This severe feature fragmentation explains why, in the Severe group, 13.0\% of Light sleep epochs were misclassified as REM (driven by high respiratory irregularity) and 9.8\% as Wake (driven by increased motor activity).

Intriguingly, the Deep sleep stage yielded counterintuitive yet physiologically sound results. The Deep sleep recognition rate in the Severe group (80.4\%) actually surpassed those of both the Mild (66.0\%) and Normal (74.7\%) groups. While the total proportion of Deep sleep is typically drastically reduced in severe OSA, the epochs where a patient successfully maintains Deep sleep often coincide with periods of relative airway stability and temporarily suppressed apnea events. Consequently, against the highly chaotic baseline of severe OSA, these rare, extremely stable signal segments form a stark feature contrast (i.e., a high signal-to-noise ratio). This allows the algorithm to anchor them with exceptional precision.

Finally, REM stage recognition remained remarkably robust across the severity spectrum, retaining an 81.7\% accuracy even in the Severe group, compared to 85.6\% in the Normal cohort. The hallmark of REM sleep is skeletal muscle atonia (minimal gross movement) coupled with active autonomic regulation (highly irregular respiration). Although REM-associated apneas are often longer and more severe, this orthogonal feature combination—drastic respiratory fluctuations alongside near-zero body movement—remains intact. The algorithm effectively capitalizes on this unique physiological signature, preventing the pathological respiratory rhythms from derailing the classification.

\subsubsection{Performance in Low Sleep Efficiency Cohorts}

To further investigate the system's clinical utility in populations with potential sleep maintenance difficulties (commonly associated with insomnia-like symptoms), we evaluated the model's performance across three subgroups stratified by sleep efficiency (SE). Although this study did not include subjective longitudinal assessments such as the Insomnia Severity Index (ISI), PSG-derived SE remains a well-established objective indicator of sleep fragmentation and maintenance impairment. Accordingly, the dataset was divided into three cohorts representing progressively reduced sleep quality: SE $\ge$ 80\%, $60\% \le$ SE $<$ 80\%, and SE $<$ 60\%, with the corresponding results summarized in Table \ref{tab:insomnia_performance}.

The Sleepal AI Lamp demonstrated remarkable robustness across all levels of sleep efficiency. Despite the substantial differences in sleep continuity among the three groups, the overall performance remained highly stable, with accuracy consistently around 76\% and Cohen's Kappa maintained near 0.65. Even in the most challenging subgroup (SE $<$ 60\%), the model achieved an accuracy of $76.3\% \pm 5.2\%$ and a Kappa coefficient of $0.651 \pm 0.079$, indicating sustained substantial agreement under highly fragmented sleep conditions.

A more detailed analysis of stage-specific performance further supports this stability. The $F_1$ score for Wake increased as SE decreased, reaching 0.862 in the SE $<$ 60\% group. This trend is physiologically consistent with the characteristics of subjects with low SE, who typically exhibit more frequent awakenings and prolonged wake periods during the night. These events generate more pronounced motion and respiratory transition patterns, which can be more reliably captured by a contactless sensing system.

In contrast, a moderate decline was observed in the classification performance of the Light and REM stages in the most severely fragmented group. This behavior is expected, as frequent micro-arousals tend to disrupt stage continuity and blur the physiological boundaries between adjacent sleep stages. Nevertheless, the recognition performance for Deep sleep remained relatively stable across the three groups, likely due to the more distinctive and rhythmic respiratory patterns associated with slow-wave sleep, which remain identifiable even in fragmented recordings.

Overall, these results demonstrate that the proposed system maintains stable sleep staging performance even in subjects with markedly reduced sleep efficiency. This robustness suggests that the Sleepal AI Lamp can serve as a reliable non-contact tool for objective sleep assessment in populations with sleep maintenance difficulties and insomnia-related phenotypes.

\begin{table}[htbp]
    \centering
    \caption{Sleep staging performance across three subgroups stratified by Sleep Efficiency (SE) (mean $\pm$ std).}
    \label{tab:insomnia_performance}
    \setlength{\tabcolsep}{12pt} 
    \begin{tabular}{lccc}
        \toprule
        Metric & SE $\ge$ 80\% & 60\% $\le$ SE $<$ 80\% & SE $<$ 60\% \\
        & ($N=156$) & ($N=82$) & ($N=22$) \\
        \midrule
        Accuracy (\%)       & 76.7 $\pm$ 5.8 & 76.4 $\pm$ 6.2 & 76.3 $\pm$ 5.2 \\
        Cohen's Kappa       & 0.668 $\pm$ 0.084 & 0.655 $\pm$ 0.095 & 0.651 $\pm$ 0.079 \\
        Macro-averaged F1   & 0.751 $\pm$ 0.071 & 0.754 $\pm$ 0.076 & 0.723 $\pm$ 0.069 \\
        \midrule
        $F_1$ Wake          & 0.841 $\pm$ 0.088 & 0.786 $\pm$ 0.130 & 0.862 $\pm$ 0.065 \\
        $F_1$ Light         & 0.725 $\pm$ 0.098 & 0.749 $\pm$ 0.079 & 0.658 $\pm$ 0.120 \\
        $F_1$ Deep          & 0.650 $\pm$ 0.185 & 0.675 $\pm$ 0.177 & 0.690 $\pm$ 0.150 \\
        $F_1$ REM           & 0.787 $\pm$ 0.132 & 0.807 $\pm$ 0.110 & 0.683 $\pm$ 0.179 \\
        \bottomrule
    \end{tabular}
\end{table}

\section{Discussion}

In this work, we presented a fully contactless, radar-based sleep staging system—the Sleepal AI Lamp—and validated its performance on a large-scale, synchronized radar-PSG dataset comprising 1,022 overnight recordings. By integrating physiologically informed features with an attention-based temporal modeling framework, the system demonstrated robust discriminative capabilities for both binary sleep-wake detection and four-class sleep architecture estimation. 

Although direct cross-dataset comparisons should be interpreted with caution due to underlying demographic differences, the proposed non-contact system compares favorably with recent validation studies of prominent consumer wearables. Existing literature\cite{robbinsAccuracyThreeCommercial2024} indicates that devices such as the Apple Watch, Oura Ring, and Fitbit generally report binary sleep-wake Kappa coefficients between 0.52 and 0.60, and four-stage Kappa values ranging from 0.55 to 0.65. In our evaluation, the Sleepal AI Lamp yielded a binary Kappa of 0.791 and a four-stage Kappa of 0.677. More importantly, while wearable validations are frequently restricted to healthy cohorts, our system maintained this competitive consistency across a substantially larger dataset that includes patients with varying severities of obstructive sleep apnea (OSA). This highlights the system's robust generalization capability and signal stability, circumventing the compliance burden and physical discomfort associated with body-worn sensors.

Beyond epoch-level accuracy, the system demonstrated high temporal fidelity in estimating sleep boundaries (sleep onset and offset), which are critical anchors for deriving clinical macro-parameters like sleep onset latency (SOL) and wake after sleep onset (WASO). The tight clustering of prediction errors around zero reflects the algorithm's capacity to reliably capture the subtle cardiopulmonary transitions between wakefulness and sleep. Furthermore, the model's resilience in the presence of sleep-disordered breathing—effectively decoupling transient OSA-induced respiratory variability from underlying sleep stage characteristics—represents a significant advancement over conventional non-contact algorithms that often degrade in pathological scenarios. This robustness is further corroborated by our analysis of low-sleep-efficiency cohorts, where the system maintained stable staging performance and exhibited enhanced sensitivity in wake detection as sleep fragmentation increased, underscoring its utility for the objective assessment of insomnia-related sleep maintenance challenges.

Despite these promising results, several limitations contextualize our findings and chart the course for future investigations. First, while the sleep laboratory environment ensured high-quality PSG ground truth, it may not perfectly mirror real-world home sleeping conditions where ambient noise, complex furniture layouts, and multiple-occupant scenarios introduce additional sensing challenges. Second, the reliance on cardiopulmonary and gross motor dynamics inherently lacks the direct cortical information provided by EEG. Consequently, distinguishing between physiologically analogous states—such as quiet wakefulness and light sleep (N1/N2)—remains the primary source of classification error. 

To address these limitations and accelerate clinical translation, future efforts will prioritize deployment in naturalistic home environments to collect longitudinal data. This will facilitate the evaluation of night-to-night reliability and user adherence. Furthermore, incorporating lightweight multimodal cues (e.g., unobtrusive bed pressure sensing or acoustic monitoring) could enrich the feature space, potentially resolving remaining staging ambiguities while preserving the fully contactless paradigm. 

\section{Conclusion}
This study validates the Sleepal AI Lamp as a highly reliable, fully contactless radar-based alternative for continuous sleep monitoring. By synergizing multi-scale cardiorespiratory and motor features with explicit temporal and chronobiological priors into a frequency-augmented deep learning architecture, the proposed framework achieves substantial agreement with gold-standard PSG. Crucially, the system transcends the limitations of many existing monitoring solutions by maintaining robust staging performance and accurate boundary estimation across a clinically diverse population, including individuals with severe obstructive sleep apnea and those exhibiting significant sleep maintenance challenges characterized by low sleep efficiency.

By eliminating the need for physical sensor attachment, this unobtrusive paradigm minimizes patient burden and preserves natural sleep behavior. Furthermore, the model's consistent performance in highly fragmented sleep scenarios highlights its potential as an objective tool for assessing insomnia-related phenotypes. Ultimately, the integration of advanced radar sensing and deep temporal modeling provides a scalable and equitable foundation for long-term sleep health management, demonstrating strong potential for routine home-based screening, remote patient monitoring, and large-scale epidemiological research. As data repository continues to expand well beyond the 1,022 nights analyzed herein, ongoing and future efforts will focus on validating the framework against this larger, richer dataset to ensure its robust applicability across increasingly diverse real-world clinical scenarios.

\bibliographystyle{IEEEtran} 
\bibliography{main} 

\end{document}